# Mechanical Properties of Graphene Papers

Yilun Liu[1,2], Bo Xie[1,2], Zhong Zhang[2,3], Quanshui Zheng[1,2] and Zhiping Xu[1,2,*]

[1]Department of Engineering Mechanics, Tsinghua University, Beijing 100084, China

[2]Center for Nano and Micro Mechanics, Tsinghua University, Beijing 100084, China

[3]National Center for Nanoscience and Technology, Beijing, 100190, China

[*]Author to whom correspondence should be addressed. E-mail: xuzp@tsinghua.edu.cn

**Abstract**

Graphene-based papers attract particular interests recently owing to their outstanding properties, the key of which is their layer-by-layer hierarchical structures similar to the biomaterials such as bone, teeth and nacre, combining intralayer strong $sp^2$ bonds and interlayer crosslinks for efficient load transfer. Here we firstly study the mechanical properties of various interlayer and intralayer crosslinks via first-principles calculations and then perform continuum model analysis for the overall mechanical properties of graphene-based papers. We find that there is a characteristic length scale $l_0$, defined as $\sqrt{Dh_0/4G}$, where $D$ is the stiffness of the graphene sheet, $h_0$ and $G$ are the height of interlayer crosslink and shear modulus respectively. When the size of the graphene sheets exceeds $3l_0$, the tension-shear (TS) chain model that are widely used for nanocomposites fails to predict the overall mechanical properties of the graphene-based papers. Instead we proposed here a deformable tension-shear (DTS) model by considering the elastic deformation of the graphene sheets, also the interlayer and intralayer crosslinks. The DTS is then applied to predict the mechanics of graphene-based paper materials under tensile loading. According to the results we thus obtain, optimal design strategies are provided for designing graphene papers with ultrahigh stiffness, strength and toughness.

## 1. Introduction

High-performance and low-cost composites are engineer's dream materials for mechanical, civil and aerospace applications. The carbon fibers, as firstly created in 1950s, are still major suppliers of high performance composites for their remarkable mechanical properties, relatively easy, cheap fabrication process and low weight. Recently, with the development of nanoscale synthesis and engineering technologies, also as inspired by the hierarchical structures in biological materials, nanocomposites featuring superior stiffness, strength and energy dissipation capacities are reported as the next-generation multifunctional super-materials (Buehler, 2006; Dunlop and Fratzl, 2010; Fratzl and Weiner, 2010; Gao et al., 2003; Ji and Gao, 2004, 2010; Kotov, 2006; Rafiee et al., 2009). For example, super-strong nanofibers such as carbon nanotubes are dispersed in polymer or metal matrices for mechanical reinforcement (Ajayan and Tour, 2007). The carbon nanotubes, with perfect cylindrical graphitic structures, attract enormous efforts towards realizing their applications in high-performance nanocomposites. However, especially for multiwalled carbon nanotubes, issues such as the limited surface-to-volume ratios due to the inaccessibility of inner walls, poor dispersion abilities in matrices due to bundle formation and also their relatively high costs prohibit wide applications as reinforcement phases.(Rafiee et al., 2010) In contrast, graphene-based nanocomposites, especially papers, impressively overcome these issues by providing two-dimensional building blocks assembled in a layer-by-layer hierarchy (Stankovich et al., 2006), which can be crosslinked by various chemicals to establish both intralayer, i.e. graphene layers are bridged on the edges, in the same plane, and interlayer load transfer (Dikin et al., 2007; Park et al., 2008; Stankovich et al., 2010; Stankovich et al., 2006; Zhu et al., 2010).

In the paper materials made from graphene and its derivatives such as graphene oxides, graphene nano-sheets as the reinforcement phase are assembled in a layer-by-layer manner (Figure 1). Because of the finite size of the graphene sheets, the in-plane tensile load can hardly be continuously transferred through intralayer bonds of the distributed graphene sheets, thus the interlayer crosslink is required to assist the tensile load transfer between adjacent layers. For graphene nanocomposites the intralayer covalent bonds are usually much stronger than interlayer crosslinks. The shear strength between adjacent graphene layers by deforming the interlayer crosslinks thus limits the load transfer between them and defines the failure mechanism of whole composites (Gong et al., 2010). The bare, van der Waals or $\pi$-orbital, interaction between graphene layers in graphite leads to a ultra-low shear strength on the order of megapascals (Yu et al., 2000). Introducing strong interlayer binding thus holds the promise in improving the shear strength. For example, nuclear irradiation creates covalent bonds bridging graphene layers (Huang et al., 2010; Telling et al., 2003), and thus enhances the energy barrier against interlayer sliding. However, this technique is difficult to control and be utilized for massive production. A more economic and flexible method, especially with controllability and reversibility, needs to be explored.

The chemical reduction method provides an efficient and cheap way to synthesis graphene sheets from exfoliated graphite by oxidation (Bai et al., 2011; Compton and Nguyen, 2010; Zhu et al., 2010). The reduced products usually contain lots of oxygen-rich chemical groups like epoxy and hydroxyl. From this point of view, the graphene oxide that features weakened in-plane mechanical properties but much improved and engineerable interlayer interactions shows great potentials in fabricating high-performance nanocomposites and papers. During the post-processing of graphene-based papers, layers of graphene or graphene oxides self-assemble in a layer-by-layer way, with additional controllability from chemical treatments. Recent experiments show that remarkable enhancement of the mechanical properties (Young's modulus, strength and toughness) could be established through introducing various functional groups, such as divalent ions and polymers (Dikin et al., 2007; Gao et al., 2011; Jeong et al., 2008; Park et al., 2008; Stankovich et al., 2010; Stankovich et al., 2007). However, a quantitative understanding of the structure-property relationship here is still lacked. Optimal design by engineering the hierarchical structures of graphene-based papers and nanocomposites is thus prohibited.

There are several existing theoretical models to treat the mechanics of composites with hierarchical structures. The shear-lag model proposed by Cox in 1952 considers the elasticity of the fiber and the interface shear between the hard and soft phases and is widely used in the fiber-reinforced composites including recently investigated carbon nanotube nanocomposites (Cox, 1952; Gao and Li, 2005). However, the graphene-based papers are assembled in a layer-by-layer pattern that is similar to biological materials such as bone, teeth and nacre. For the mechanical properties of biological materials with hierarchical microstructures, Gao and his collaborators (Gao et al., 2003; Ji and Gao, 2004) propose the tension-shear (TS) chain model. This model is further extended to capture the material failure mechanisms and distribution of stiff-phase platelets in those biological materials (Barthelat et al., 2007; Tang et al., 2007; Zhang et al., 2010). In the tension-shear chain model the mineral bones is considered as rigid bars or platelets, so the shear stress at the interface between the mineral and the protein is uniform. In contrast, this assumption fails for graphene-based papers due to their extremely large aspect ratios. The in-plane dimension of a graphene sheet is on the order of micrometers and its thickness is less than 1 nm, the elastic deformation of the graphene is thus comparable to shear deformation of the interlayer crosslinks. As we will see later in the text, the elasticity of graphene must be considered in the continuum model to successfully predict overall mechanical properties of the graphene-based papers.

There are three key factors that define the overall performance of graphene oxide based nanocomposites: (1) intralayer mechanical properties, defined by $sp^2$ carbon-carbon covalent bonds and crosslinks at graphene edges, (2) mechanical properties of interlayer crosslinks and (3) structural characteristics of the nanocomposites, i.e. the size, and spatial position distribution of graphene sheets and their crosslinks. Based on the knowledge from previous experimental studies on the structure and mechanical property of graphene oxide nanocomposites, we here firstly quantify the mechanical properties of various crosslink mechanisms, including both intralayer and interlayer interactions. Based on these arguments, in the spirits of shear-lag and tension-shear chain models, we here develop the deformable tension-shear (DTS) model to capture the mechanical properties of graphene papers with interlayer, intralayer crosslinks and structural hierarchy as mentioned before. The overall mechanical performance of the nanocomposites is predicted through a multiscale approach with parameters fed by first-principle calculations.

This paper is organized as follows. In Section 2, after describing the microstructures of graphene-based papers (Section 2.1), we firstly introduce the details of our first-principles calculations (Section 2.2), then the mechanics of both interlayer and intralayer crosslinks (Section 2.3 and 2.4). The DTS model is developed in Section 3.1 and the importance of considering intralayer deformation in graphene sheets in this model is introduced in Section 3.2. Then we applied this model to investigate the overall mechanical properties of graphene-based papers by considering both interlayer and intralayer crosslinks (Section 3.3 and 3.4). Discussions on these results are presented in Section 4 before concluding in Section 5, focusing on the failure mechanism (Section 4.1), optimal design strategies (Section 4.2) and some additional comments (Section 4.3).

## 2. Mechanical Properties of Interlayer and Intralayer Crosslinks

### 2.1 The structure of Graphene-based Papers

A schematic plot for the analytic model of graphene-based papers is shown in Figure 1(a). Graphene or graphene oxide sheets are highly ordered structures where graphene sheets are stacked layer by layer through both intralayer and interlayer crosslinks. This order is of critical importance in passing the nanoscale interlayer interactions to their macroscopic mechanical properties (Ci et al., 2008). By assuming that all the graphene sheets are staggered as shown in Figure 1(a), and have the same size, an interlinked elastic plate model can be constructed. A two-dimensional representative volume element (RVE) is used to represent the whole composites, where one sheet overlaps with its two neighbors by half of its length. The whole nanocomposite can be constructed by repeatedly stacking the RVE cell. As the tensile force is applied to the graphene-based paper, the tensile load is mainly sustained by the monatomic graphene sheet. Between the adjacent graphene sheets the in-plane tensile force is transferred by both the interlayer shear and intralayer tension. Thus the mechanical properties of the graphene-based paper are determined by the mechanical properties of interlayer and intralayer crosslinks, the mechanical properties and size of the graphene oxide sheets. However the mechanical properties of the graphene interlayer and intralayer crosslinks have not been studied yet. So in order to predict the mechanical properties of the graphene-based materials we must firstly understand the mechanics of various interlayer and intralayer crosslinks, via first principles calculations here.

### 2.2 First-Principles Calculations for Mechanical Properties of Crosslinks

The mechanics of crosslinks between adjacent graphene sheets lying in parallel is captured by a supercell approach as shown in Figure 1(a). The interfaces between graphene layers with various functional groups are investigated through a rhombic supercell of 1.72 nm×1.72 nm. A vacuum layer of 2 nm is used in the direction normal to interface, representing the isolated boundary condition for graphene bilayers. The structure and mechanical properties of this hybrid system are subsequently investigated using plane-wave basis sets based density functional theory (DFT) methods. The generalized gradient approximation (GGA) in Perdew-Burke-Ernzerhof (PBE) parameter settings is used for the exchange-correlation functional and projector augmented wave (PAW) potentials are used for ion-electron interaction. We use the Vienna Ab-inito Simulation Package (VASP) for the calculations. For all results presented, an energy cut-off of 300 eV is used for plane-wave basis sets and single gamma point is used for Brillouin zone integration as we have a large supercell. These settings have been verified to achieve a total energy convergence less than 1 meV/atom. For geometry relaxation, the force on atoms is converged within 0.01 eV/Å. All structures are initially optimized using a conjugated gradient method. It should be noticed that in our simulation a supercell is used, thus the model here represents actually graphene papers with a periodic crosslink array. The density is one crosslinker per 2.562 nm$^2$.

The key parameters here for the crosslinks are the mechanical properties of intralayer crosslink under tensile loads and interlayer crosslink under shear loads. Tensile behavior of intralayer crosslinks is investigated by directly applying a tensile in-plane load on the supercell and geometrical optimization afterwards. The stress is calculated by assuming the graphene sheet as a thin shell with a thickness of 1 nm. The performance of interlayer crosslinks is measured by vertically (transversely, in <1-100> direction) moving one layer of graphene sheet with respect to its neighbor step by step with an interval of 0.01 nm. After shear displacement is applied, the degrees of freedom of all carbon atoms in the shear direction are fixed while the other two directions are set to be freely relaxed for subsequent geometrical optimization calculations. The interlayer stress is calculated by summing up all forces acting on the carbon atoms of one graphene layer in the shear direction, and then being divided by the area of supercell in use.

### 2.3 Interlayer Crosslink Chemistry and Mechanics

The values of shear strength $\tau_s$ between adjacent graphene sheets, with bare van der Waals (or $\pi$-orbital) interactions or a variety of crosslink mechanisms as calculated by the DFT calculations, are summarized in Figure 1(b). The value $\tau_s$ for bare interface between graphene sheets is 367 MPa (along the <1-100> direction), which is although in consistence with earlier studies (Bichoutskaia et al., 2009; Bichoutskaia et al., 2006; Xu et al., 2008), but two orders higher than experimental measurements 0.48 MPa (Bichoutskaia et al., 2009; Yu et al., 2000). It is noticed that in the DFT-based calculations, the van der Waals nature cannot be captured, while the interlayer interaction is described as electronic coupling between $\pi$-orbitals at the ground state (Bichoutskaia et al., 2006). Additionally, the shear stress is calculated along the high-symmetry direction and at zero temperature, which result in the higher calculated value here in comparison to the experimental measurements. Even though, this calculated value is still three orders lower than the theoretical in-plane strength (~ 100 GPa), which explains why the interlayer load transfer is the bottleneck for overall performance of graphitic materials.

A variety of functionalization types are introduced to the graphene sheets during the oxidation, reduction and post-processing treatments. Under hydrated condition, the hydrogen-bond (Hbond) network forming between water molecule or epoxy groups on the graphene sheet will expand the interlayer distance from 0.335 nm in graphite, lowering the interlayer interaction and breaking the A-B stacking registry between adjacent graphene sheets (Medhekar et al., 2010). The shear strength can be further lowered for a thicker water film, as structured water is available only for few layers adjacent to the graphene sheets (Xu et al., 2010).

In our calculation, we find firstly that a single water molecule between the graphene layers lowers the strength to 26 MPa and expands the interlayer distance to 0.6 nm. Two epoxy groups on graphene sheets, facing oppositely, lead to a slightly enhanced $\tau_s = 30$ MPa and interlayer distance $h_0 = 0.5$ nm, in the absence of interstitial water molecules. For graphene oxides, surface chemical groups such as epoxy and hydroxyl can assist the formation of interlayer Hbond networks. Two types of hydrogen bonds are considered in this work (Medhekar et al., 2010). the one formed between an epoxy group and a hydroxyl group on another graphene sheet (HB1) leads to a shear strength of 103 MPa. While the hydrogen bond formed between two hydroxyl groups (HB2) has a shear strength $\tau_s$ of 88 MPa. The interlayer distances for both structures are 0.55 nm.

Impressively, the coordinative bonds introduced by divalent atoms or ions such as intercalating magnesium atoms improve tremendously the interlayer shear strength (Liu et al., 2011; Park et al., 2008). The coordination chemistry, where ligands provide electron pairs to form bonds with metal atoms or ions, has comparable mechanical performance with covalent bonds. In our calculations, the alkoxide and dative bonds forming by intercalating magnesium atoms are investigated. The structures are shown in Figure 1(a). The results show that the strong coordinative bonding (CB) between magnesium and the oxygen atoms directly enhances the interlayer shear strength to 811 MPa, which is defined by bond breaking between the magnesium and oxygen atoms that are attached to the graphene sheet. The DFT results for $\tau_s$ and $d$ are summarized in Figure 1(b).

### 2.4 Intralayer Crosslinks

In addition to the interlayer crosslinks connecting basal planes of graphene sheets, alkaline earth metals can also strongly bind the graphene sheets at their edges through coordinative bonds, as a carboxylic acid group. Two oxygen atoms are bridging the metal atoms from two sides, to the carbon atoms connected to the zigzag edges of graphene sheets (Park et al., 2008). Our DFT calculation results show that for a carboxylic acid group connecting graphene sheets with an edge density of one crosslinker per 2.71 nm, the tensile stiffness and strength are 57.4 and 2.85 GPa respectively, by considering the thickness of graphene sheet to be 1 nm (Liu et al., 2011), where the failure is defined by the bond breaking between the magnesium atom and oxygen atom in the carboxylic acid group. During tension of the graphene-based paper, the distance between the edges of the in-plane adjacent graphene sheets increases, so that part of the tensile load in the graphene sheet can be transferred by the intralayer crosslinks.

### 3. The Deformable Tension-Shear (DTS) Model

### 3.1 The Model

Since the hierarchical structure of the graphene-based paper is similar to the biomaterials of nacre, here we propose a continuum model based on the tension-shear model to predict the mechanical properties of the graphene-based papers, as illustrated in Figure 2(a). In this model we consider the elastic deformation of the graphene sheet during tension that is missing in the tension-shear model, so name it the deformable tension-shear model. As the illustration of Figure 2(a) and (b) we consider the interlayer crosslink as a continuum media and the intralayer crosslink is simplified as a spring connecting the edges of the in-plane adjacent graphene sheets. In the continuum model the geometry of the composite is characterized through the graphene sheet size $l$, interlayer distance $h_0$. The stiffness of the graphene sheets is defined as $D = Yh$, where $Y$ is the Young's modulus, to avoid the definition of graphene sheet thickness $h$ (Huang et al., 2006; Wang et al., 2005). Only deformation in the tensile (in-plane) direction is considered. In a linearly elastic approximation, the mechanical resistance to interlayer shear load between adjacent graphene layers is defined by interlayer shear stiffness $G$ and critical shear strain $\gamma_{cr}$ (maximal shear strain in prior to the failure of the interlayer crosslinks). We use the spring constant $k$ per unit width to character the effective stiffness of intralayer crosslink.

As a tensile force is applied on the overall structure, the tensile loads act on RVE cell is $F_0$ and $F_k$ (Figure 2(a) and (b)), where $F_0$ denotes the tensile force per unit width at $x = l/2$ where has no shear contribution, and $F_k = k\Delta r$ denotes the force per unit width between two graphene sheets connected by intralayer crosslinks with extension $\Delta r$, and $F(x)$ ($0 < x < l$) is used to denote the tensile force distribution in the graphene sheet. The in-plane displacements in the graphene sheet 1 and 2 are expressed as $u_1(x)$ and $u_2(x)$ respectively where $x$ is the position in the graphene sheet. Consequently, the shear stress in the interlayer continuum at position $x$ is $\tau(x) = G[u_1(x) - u_2(x)]/h_0$.

For graphene sheets 1 and 2 in one RVE (Figure 2(b)), the equilibrium equations are

$$D\frac{\partial^2 u_1(x)}{\partial x^2} = 2G\gamma_1(x),$$
$$D\frac{\partial^2 u_2(x)}{\partial x^2} = 2G\gamma_2(x). \tag{1a}$$

As the shear strain in the interlayer continuums are $\gamma_1(x) = [u_1(x) - u_2(x)]/h_0$ and $\gamma_2(x) = [u_2(x) - u_1(x)]/h_0$, Eqs. (1a) can be rewritten as

$$D\frac{\partial^2 u_1(x)}{\partial x^2} = 2G\frac{u_1(x) - u_2(x)}{h_0},$$
$$D\frac{\partial^2 u_2(x)}{\partial x^2} = 2G\frac{u_2(x) - u_1(x)}{h_0}. \tag{1b}$$

By solving these equations we obtain the general solutions of Eqs. (1b) as

$$u_1(x) + u_2(x) = A_1 + A_2 x,$$
$$u_1(x) - u_2(x) = A_3 \sinh\frac{x}{l_0} + A_4 \cosh\frac{x}{l_0}, \tag{2}$$

with four undetermined parameters $A_1$, $A_2$, $A_3$, $A_4$, which can be determined through introducing boundary conditions of graphene sheets 1 and 2 at $x = 0$ and $x = l$, where $l_0 = \sqrt{\dfrac{Dh_0}{4G}}$ is a length parameter determined by the graphene sheet stiffness and interlayer properties. For graphene sheet 1, at the left edge, i.e. $x = 0$, the tensile force in the graphene sheet equals to the force in the intralayer crosslink spring

$$D\frac{\partial u_1(0)}{\partial x} = k\Delta r_1, \tag{3a}$$

where $\Delta r_1$ is the extension of the intralyer crosslink and $k$ is the spring constant. For graphene sheet 2, the tensile force is $F_0$

$$D\frac{\partial u_2(0)}{\partial x} = F_0. \tag{3b}$$

Similarly, at the right edge, i.e. $x = l$, the tensile forces in graphene sheet 1 and 2 are

$$D\frac{\partial u_1(l)}{\partial x} = F_0, \quad D\frac{\partial u_2(l)}{\partial x} = k\Delta r_2, \tag{3c}$$

where $\Delta r_2$ is the extension of intralayer crosslink at the right edge of graphene sheet 2. The extension of intralayer crosslinks can be determined by the displacements of neighboring graphene sheets, i.e. $\Delta r_1 = u_1(0) - u_2'(0)$ (Figure 2(c)) based on the assumption that the deformation in repeated RVEs is the same and $u_1'(0) - u_2'(0) = u_1(l) - u_2(l)$. The superscript ' and '' represent the two repeated constructing cells of graphen sheet 1 and 2. In addition, the continuous condition leads to $u_1'(0) = u_2(0)$ and we have $\Delta r_1 = \Delta r_2 = u_1(0) - u_2(0) + u_1(l) - u_2(l)$.

### 3.2 The Impacts of Intralayer Deformation in the Graphene Sheet

Firstly we discuss the impacts of the elastic deformation of graphene sheets to the mechanical properties of graphene-based paper. Thus here we neglect the intralayer crosslink that will be discussed later, i.e. by letting $k = 0$ or $F_k = 0$. By letting the general solution (2) equal to the boundary equations (3) at $x = 0$ and $x = l$, we obtain the four undetermined parameters, i.e. $A_2 = \dfrac{F_0}{D}$, $A_3 = \dfrac{F_0 l_0}{D}$, $A_4 = \dfrac{F_0 l_0 (1 + \cosh l / l_0)}{D \sinh l / l_0}$ and $A_1$ is the rigid body displacement no influence to the elastic deformation of RVE. The tensile force is calculated as $F = D\dfrac{\partial u_1}{\partial x}$, and the tensile force distribution in one graphene sheet ($0 < x < l$) is

$$F = \frac{F_0}{2}[1 - \cosh(\frac{x}{l_0}) + \frac{1+c}{s}\sinh(\frac{x}{l_0})], \tag{4}$$

where $s = \sinh(l/l_0)$ and $c = \cosh(l/l_0)$. The interlayer shear strain is $\gamma_1(x) = [u_1(x) - u_2(x)]/h_0$, correspondingly, the shear strain distribution in the interlayer crosslink is

$$\gamma_1(x) = \frac{F_0 l_0}{D h_0}[(1+c)\cosh(\frac{x}{l_0}) - s\sinh(\frac{x}{l_0})] / s, \tag{5}$$

For graphene sheets with different lengths as $l/l_0 = 0.5, 1.0, 2.0, 3.0$ and $5.0$, we plot the distribution of in-plane tensile force and interlayer shear strain along the length direction in Figure 3(a) and (b), respectively. We can clearly see that at the open edge ($x = 0$), the tensile force is zero and at the midpoint of graphene sheet ($x = l$), $F = F_0$. Additionally, as $l$ is small ($l/l_0 < 2$), the force distribution function is linear along the graphene sheet profile which is similar to the tension-shear chain model prediction. While as $l/l_0$ reaches 5, the linearity is only kept well when close to the edge of graphene sheets, i.e. $x$ is close to 0 or $l$. In contrast, the central part of the force distribution profile becomes flat, suggesting reduced load transfer through interlayer crosslinks. This is also evidenced by the small shear strain amplitude in central part ($x \sim l/2$) of the crosslinks as plotted in Figure 3(b). Moreover, the maximal shear strain is reached at the edges of the graphene sheet, suggesting that the failure of interlayer binding will be initialized there.

If the elastic deformation of the graphene sheets is not accounted for, i.e. the shear strain distribution along the graphene sheets is uniform, for the given force $F_0$ that is completely transferred by the two adjacent interlayer shear (Figure 1(b)) we obtain the shear strain as $\gamma_{\text{rigid}} = F_0/(2lG)$. In order to compare with Eq. (5) according to the definition of $l_0 = \sqrt{\dfrac{Dh_0}{4G}}$, the shear strain is written as

$$\gamma_{\text{rigid}} = \frac{F_0 l_0}{D h_0} \frac{2 l_0}{l} . \tag{6}$$

In Eq. (5) it is noticed that as the ratio $l/l_0$ approaches 0, the term $[(1+c)\cosh(\frac{x}{l_0}) - s\sinh(\frac{x}{l_0})]/s$ $(0 < x < l)$ converges to a constant of $2l_0/l$, so that the Eq. (5) will degenerate into Eq. (6) and the DTS model is reduced into the commonly used TS model. This situation could be arrived when the stiffness of interlayer crosslink is much lower than that of the graphene sheet, or the size of graphene sheet is much smaller in comparison to $l_0$.

For comparison, we further calculate the ratio between the maximum shear strain $\gamma_{\text{max}}$ in the DTS model and $\gamma_{\text{rigid}}$ in the rigid approximation for graphene sheets with different lengths

$$\frac{\gamma_{\text{max}}}{\gamma_{\text{rigid}}} = \frac{l(1+c)}{2 l_0 s} . \tag{7}$$

As shown in the inset of Figure 3(b), we find from Eq. (7) that as the size of the graphene sheet increases, this ratio increases almost linearly, that means the elastic deformation of graphene has more impacts on the overall mechanical properties of the whole material. As larger graphene sheets are able to stabilize the paper structure and establish a uniform crosslink pattern, which further enhance the overall mechanical properties of graphene-based papers, we must consider this elastic deformation in developing a continuum model to predict these properties.

The parameter $l_0$ characterizes a typical length scale for load transfer between adjacent graphene sheets through interlayer crosslinks. As shown in Figure 3, the tensile load in the graphene sheets are mostly transferred by the interlayer crosslink within the length of $l_0$ from the edges of the graphene sheets. As the maximum shear strain of the interlayer crosslink is reached at the edges of the graphene sheets, the strength of the graphene-based papers is reached when the interlayer shear strain reaches the critical shear strain $\gamma_{\text{cr}}$ of the interlayer crosslink at the edges. Consequently, let $x = 0$ or $x = l$ in Eq. (5) we obtain the relationship between the critical shear strain and the tensile force as

$$\gamma_{\text{cr}} = \frac{F_0 l_0}{D h_0} \frac{1+c}{s} . \tag{8}$$

The maximal effective tensile stress $F_0/2h_0$, or strength $\sigma_s$ in the RVE, is predicted as

$$\sigma_s = \frac{s \gamma_{\text{cr}} D}{2(1+c) l_0}. \tag{9}$$

For the RVEs constructed by graphene sheets with different lengths, we plot their strength in Figure 4(a), which shows that as $l$ reaches $5 l_0$, the strength value of the graphene paper converges to a constant value $\gamma_{\text{cr}} D/(2 l_0)$ . By substituting with $l_0 = \sqrt{\dfrac{D h_0}{4G}}$ , the upper bound strength can be written as $\sigma_s = \dfrac{2 \gamma_{\text{cr}} G l_0}{h_0}$ . In the tension-shear chain model, when the size of the rigid platelet is $2 l_0$, the effective tensile strength of RVE is also estimated as $\dfrac{2 \gamma_{\text{cr}} G l_0}{h_0}$ when the whole interlayer materials reaches the same critical strain $\gamma_{\text{cr}}$. Because the tensile load transferred by the interlayer shear deformation locates mainly within a distance of $l_0$ from the graphene edges, so the tensile strength is mainly contributed by the interlayer crosslinks in these regions. This is the reason why the upper bound tensile strength of RVE constructed by very large size graphene sheets is equivalent to the tensile strength of the RVE constructed by a rigid platelet with a length of $2l_0$ and the same interlayer crosslink density. According to our first-principles calculations, we calculate the typical length scale $l_0$ for the magnesium coordinative crosslink is about 6.8 nm. While in graphene-based papers as synthesized by current techniques, the size of graphene sheet is usually on the order from micrometers to millimeters, so the elastic deformation of the graphene sheets has a strong impact on the overall mechanical properties and must be accounted in order to make accurate predictions.

### 3.3 Effective Young's Modulus and Toughness

We can also define an effective Young's modulus $Y_{eff}$ for the whole structure as $(F_0/2h_0)/(\Delta/l)$, where $F_0/2h_0$ is the effective tensile stress and $\Delta = [u_1(l) - u_2(0)]$ is the elongation of the RVE cell. The effective Young's modulus is

$$Y_{eff} = \frac{D}{2h_0} \frac{1}{\frac{1}{2} + \frac{1+c}{s}\frac{l_0}{l}} \ . \tag{10}$$

Eq. (10) shows that $Y_{eff}$ scales linearly with $l$ as it is small and converges to a constant $Y_{eff} = D/h_0$ for large graphene sheets ($l > 10l_0$), contributed solely by the tensile deformation in the graphene sheets, as shown in Figure 4(a).

Toughness is the key material property to describe its ability to absorb energy and deform without fracturing. Here in our graphene paper model, the toughness can be divided into two parts, i.e. the strain energy stored in the in-plane tensile deformation of graphene sheet and the one in the shear deformation of the interlayer crosslinks. According to the tensile force and the shear strain distribution as obtained in previous sections, we now can directly calculate the tensile and shear energy in RVE as $2\int_0^l \frac{F^2}{2D}\mathrm{d}x$ and $2\int_0^l G\gamma^2 h_0 \mathrm{d}x$, respectively. So the total strain energy density stored in both the graphene sheets and the interlayer crosslinks before material failure, is calculated as

$$T = sh_0\gamma_{cr}^2 D \frac{\frac{1}{2}s + (1+c)\frac{l_0}{l}}{4(1+c)^2 l_0^2}, \tag{11}$$

which has a maximum value of $0.178\gamma_{cr}^2 Dh_0/l_0^2$ at $l = 3.3\ l_0$. For graphene sheets with different lengths we plot the strain density of intralayer tensile contribution, interlayer shear contribution and their sum separately in Figure 4(b). The results indicate that while the shear contribution decays as the graphene sheet size $l$ increases and becomes almost zero as $l > 10l_0$, the tensile contribution has a maximum at $l \sim 5l_0$ and converges to a dominating contribution at large $l$ values.

By combining Eq. (9) through (11) we conclude that the effective Young's modulus and tensile strength can be enhanced by increasing the graphene sheet size, and there exists an optimal graphene sheet size for maximal toughness.

### 3.4 Intralayer Crosslinks

In addition to the interlayer crosslinks that connect basal planes of graphene sheets, we also investigate the effects of intralayer crosslinks on the overall mechanical properties of the graphene-based papers in our continuum model. As an example, the edges of the adjacent graphene sheets can be crosslinked by alkaline earth metals through coordinative bonds, as a carboxylic acid group. When the intralayer crosslinks are taken into account, we introduce $k$ as the effective spring constant of unit width for the intralayer crosslinks. By substituting the general solution (2) into the boundary equations (3), we solve the four undetermined parameters as

$$A_2 = \frac{2F_0}{D} - \frac{F_0}{l_0} / (\frac{D}{l_0} + 2k\frac{1+c}{s}),$$

$$A_3 = -F_0 / (\frac{D}{l_0} + 2k\frac{1+c}{s}),$$

$$A_4 = \frac{1+c}{s}F_0 / (\frac{D}{l_0} + 2k\frac{1+c}{s}).$$

(12)

Similar to our previous analysis for the graphene-based papers without intralayer crosslinks, the parameter $A_1$ represents the rigid body displacement and can be an arbitrary value. As the spring constant $k$ approaches 0, $A_2$, $A_3$, $A_4$ degenerate back to the values of the solution without intralayer crosslinks, and we obtain the interlayer shear strain distribution as

$$\gamma = \frac{A_3 \sinh x / l_0 + A_4 \cosh x / l_0}{h_0}.$$

(13)

We further define $k_l$ as $kl_0/D$, the ratio between the stiffness of intralayer crosslink and the graphene sheet. For different $k_l$ values 0, 0.1, 0.4 (corresponding to the magnesium coordinative intralayer crosslink of our DFT calculation) and 1.0, we plot the interlayer shear strain distribution along the graphene sheet in Figure 5(a). The maximum shear strain occurs at the edges of the graphene sheets. But it can be seen that at the same tensile force $F_0$, the shear strain distribution in the interlayer crosslink is significantly reduced and the intralayer crosslink elasticity contribution to the in-plane load transfer increases. Then based on the analytical solutions of the displacements in the RVE under tension we obtain the effective tensile modulus and strength accounting for the influence of both interlayer and intralayer crosslinks. Because of the intralayer crosslink, the effective tensile force acting on the RVE is $F_0 + F_k$ and also the tensile stress is $\sigma = (F_0 + F_k)/2h_0$, where $F_k$ is contributed by the intralayer crosslink. The effective tensile strain of the RVE is $\varepsilon = [u_1(l) - u_2(0)]/l$. We thus obtain the effective tensile modulus $\sigma/\varepsilon$ as

$$Y_{\text{eff}} = \frac{D}{h_0} \frac{\frac{1}{2} + k_l \frac{1+c}{s} \frac{1}{1 + 2k_l \frac{1+c}{s}}}{1 - \frac{1}{2} \frac{1}{1 + 2k_l \frac{1+c}{s}} + \frac{l_0}{l} \frac{1+c}{s} \frac{1}{1 + 2k_l \frac{1+c}{s}}} . \tag{14}$$

For graphene sheets with different lengths, i.e. $l/l_0 = 0.5, 1, 3, 5$, we plot the relationship between the effective tensile modulus $Y_{\text{eff}}$ and the intralayer crosslink stiffness $k_l$ in Figure 5(b). As the stiffness of intralayer crosslinks increases, the contribution of the in-plane tension of the graphene sheets also increases, thus the effective tensile modulus of the RVE is enhanced. However, $Y_{\text{eff}}$ converges to $D/h_0$, the same value as the graphene papers without intralayer crosslinks, whenever $k_l$ and $l$ increase.

As the maximum shear strain occurs at the edges of the graphene sheets, the failure firstly occurs at interlayer crosslinks of the edge or intralayer crosslinks. Here to obtain some insights of the impacts from intralayer crosslinks to the strength of the graphene-based papers we only consider the failure of the interlayer crosslink. By equating the interlayer shear strain of edges with the critical strain of the interlayer crosslink, we obtain the maximum allowable tensile force $F_0$ is

$$F_0 = F_{\text{normal}} + \gamma_{\text{cr}} \frac{h_0 D}{l_0} 2k_l , \tag{15}$$

where $F_{\text{normal}}$ is the maximum $F_0$ without intralayer crosslinks. The other in-plane tensile force $F_k$ loaded by intralayer crosslink is

$$F_k = \gamma_{\text{cr}} \frac{h_0 D}{l_0} 2k_l . \tag{16}$$

Finally we can obtain the strength of RVE accounting for both the intralayer and interlayer crosslink as

$$\sigma_{\text{cr}} = \sigma_{\text{normal}} + 2k_l \frac{D\gamma_{\text{cr}}}{l_0} , \tag{17}$$

where $\sigma_{\text{normal}}$ stands for the strength without intralayer crosslinks. From Eq. (17) we find that the influence of the intralayer crosslink to the strength is $2k_l D\gamma_{\text{cr}}/l_0$, that is proportional to the stiffness of the stiffness $k_l$ of intralayer crosslinks and independent on the graphene sheet size. In brief, the intralayer crosslink can enhance both the modulus and strength of the graphene-based papers, where the enhancement effect of the modulus improves for graphene sheets of small size, while for the strength there is no such size effects.

## 4. Discussion

### 4.1 Failure Mechanisms

When the applied load exceeds the strength, the whole composite starts to fail by breaking at the weakest point, which could be at either the intralayer or interlayer crosslinks. For the first situation, the failure is defined by that the displacement of intralayer crosslink exceeds $\Delta r_{\text{cr}}$. and for the second mechanism, it is defined by the critical shear strain $\gamma_{\text{cr}}$ of interlayer crosslinks. Correspondingly in our analytical model, if $\Delta r_{\text{cr}} > 2h\gamma_{\text{cr}}$, as in the situation where magnesium-based coordinative bonding crosslinks in-plane graphene sheets, the failure of the structure starts at the interlayer crosslinks, initialized from the edge of the graphene sheets where the shear strain distribution reaches its maximum (Figure 3(b) and Figure 5). On the other hand, if the intralayer crosslinks break firstly and the elastic energy stored there is dissipated. As the second step, the load is then sustained by interlayer crosslinks only, where material failure also starts from the edge of graphene sheets.

The failure of the interlayer crosslinks firstly occurs at the edges of the graphene sheets. Then as the failure develops, the length of the interlayer crosslinks that are bearing loads decreases. However the effective load transfer length is about $l_0$, so that the strength reduction of the graphene-based papers is negligible if the remaining length is larger than $3l_0$ (Figure 4). Because of this property the graphene-base papers can also bear the load after the failure starts. This material robustness is of key importance in applications.

### 4.2 Optimal Design of the Mechanical Properties

By substituting all the mechanical parameters obtained from first-principles calculations to Eq. (3)-(5), we plot the strength and toughness of three representative crosslink types in Figure 6(a). For graphite, where the interlayer distance between graphene sheets is $h_0 = 0.335$ nm, interlayer shear modulus $G = 2.548$ GPa and maximum shear strain $\gamma_{cr} = 0.144$, the maximal strength and toughness are 6.3 GPa and 38 MPa respectively. With the enhancement from Mg-centered coordinative bonds, $G = 970$ MPa, $h_0 = 0.71$ nm and $\gamma_{cr} = 0.76$, thus we have the shear strength approaching 14 GPa, and toughness of 400 MPa. While with the Hbond networks formed between epoxy and hydroxyl groups, where $G = 763$ MPa, $h_0 = 0.545$ nm and $\gamma_{cr} = 0.135$, the shear strength and toughness are 2.5 GPa and 10 MPa. It is also noticeable that for all the structures, the toughness maximizes for the graphene size at $l \sim 23$ nm. Furthermore we vary the shear modulus from 970 MPa to 97 and 9.7 MPa, and graphene sheet size from 100 nm to 1 nm to track their impacts on mechanical properties of the graphene papers. Figure 6(b) shows that the strength changes from 10 MPa to 10 GPa and toughness changes from 3 MPa to 400 MPa. Thus by increasing the graphene sheet size and crosslink strength, the strength and toughness of the materials will be enhanced cooperatively.

Graphene-based paper, and in general nanocomposites, are projected for various applications requiring ultrastrong, lightweight and multifunctional features. One additional benefit from interlayer crosslinks is that as they are introduced, the interlayer distance is expanded. In comparison to metals such as aluminum with a mass density 2.7 $g/cm^3$, the density of graphite, 2.25 $g/cm^3$, is already lower. For the coordinative bonds as an example, the expansion due to coordinative bond will further lower the density approximately by half.

### 4.3 Additional Comments on the Crosslink Mechanisms

In order to improve the overall properties (stiffness, strength and toughness) of graphene-based nanocomposites, three requirements must be fulfilled. (1) As the graphene sheets usually have finite size of micrometers, an effective way to increase intralayer integration is needed to transfer the in-plane tensile load. (2) Due to the randomness in distribution of graphene sheet sizes, positions and stacking orders, the load transfer between adjacent graphene layers must be enhanced by introducing interlayer bridging. (3) The crosslink between graphene sheets needs to be able to re-associate after breaking under tensile or shear loads. The intralayer and interlayer crosslinks ensures a high toughness by utilizing the concept of 'sacrificial bonds' in biological materials such as bones (Ritchie et al., 2009). Especially the interlayer crosslinks can form again after breaking and relative sliding between adjacent graphene layers. One interesting observation in Park et al.'s report (Park et al., 2008) is that repeated cyclic loading of a metal-modified paper sample with small force and under slow rate may allow for chemically annealing to the best crosslink structure. Thus the formation and adjustment of crosslinks under loads, which could lead to ordering of the hierarchical structures, are of critical importance for the mechanical properties enhancements of these nanocomposites. To fulfill requirements (1) and (2), strong bonding like covalent and coordinative bonds take the advantage. However for (3), the covalent bond, that is short-ranged and directional, is difficult to form again after breaking. The coordinative bonds could play a critical role in the scenario with the presence of metal ions, epoxy and hydroxyl groups on graphene oxide sheets (Liu et al., 2011). On the other hand, although the hydrogen bonds and van der Waals forces are much weaker, they could provide remarkable mechanical contributions as working cooperatively.

## 5. Conclusions

In summary, the mechanics of graphene-based papers where graphene sheets are assembled in a layer-by-layer manner is studied here by first-principle calculations and the deformable tension-shear model. In comparison to the tension-shear chain model, we further include the elastic deformation of the graphene sheets that has key impacts to the overall mechanical properties of the graphene-based papers, as evidence here by the calculations and discussion. This model can also be extended to other paper-, fiber- or bundle-like materials such as those of carbon nanotubes. According to the analysis results obtained from the DTS model, we find that by simultaneously tuning the graphene sheet size and crosslink mechanisms, an optimal design of graphene-based papers can be established with well-appreciated mechanical performance. This design concept is reminiscent of many nature materials such as bones and mollusk shells (Fratzl et al., 2004; Ji and Gao, 2004), where hierarchical structures are utilized and optimized for billions of years to fit requirements in enabling biological functions. While in engineering for best mechanical performance, we here can simply pick the strongest two-dimensional materials – graphene as show here and further engineer the crosslink mechanisms between them, towards a supermaterial.

The results and model presented in this work directly map the atomistic level interactions into macroscopic mechanical performance, and thus offer a rational design strategy for high-performance nanocomposites. To characterize structural complexities such as the random distribution of graphene sheet sizes, positions, stacking orders, crosslink density and strength, and quantify their thermal and electrical properties, need to be explored further (Hartmann and Fratzl, 2009; Zhang et al., 2010).

### Acknowledgements

This work is supported by Tsinghua University through the Key Talent Support Program, and the National Science Foundation of China through Young Scholar Grant 11002079 (ZX) and Key Program Grant 10832005, the China 973 Program No. 2007CB936803, 863 Program No. 2008AA03Z302 (QZ). This work is also supported by Shanghai Supercomputer Center of China.

**Figures and Captions**

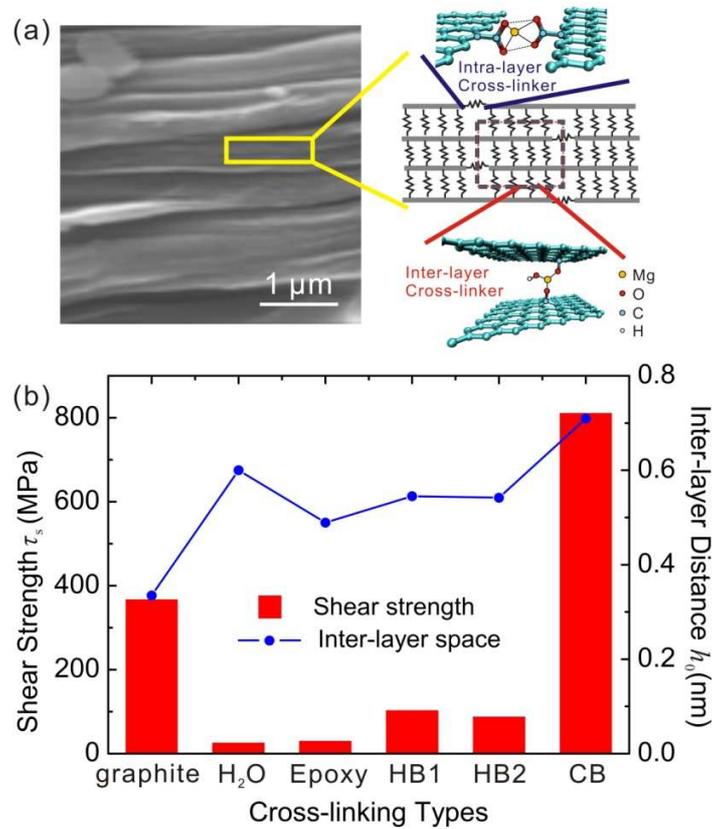

**Figure 1.** (a) A scanning electron microscopy (SEM) image of graphene oxide papers and an analytical model showing the layered structures of graphene sheets, the intralayer and interlayer crosslinks. An atomic representation of the bridging structure is also shown. (b) Shear strength and interlayer distance values between adjacent graphene sheets that are bridged through several crosslink types, including bare interactions in graphite, interstitial water layers ($H_2O$), two epoxy groups (Epoxy), hydrogen bond networks formed between epoxy and hydroxyl groups (HB1), two hydroxyl groups (HB2) and divalent atom assisted coordinative bonds (CB).

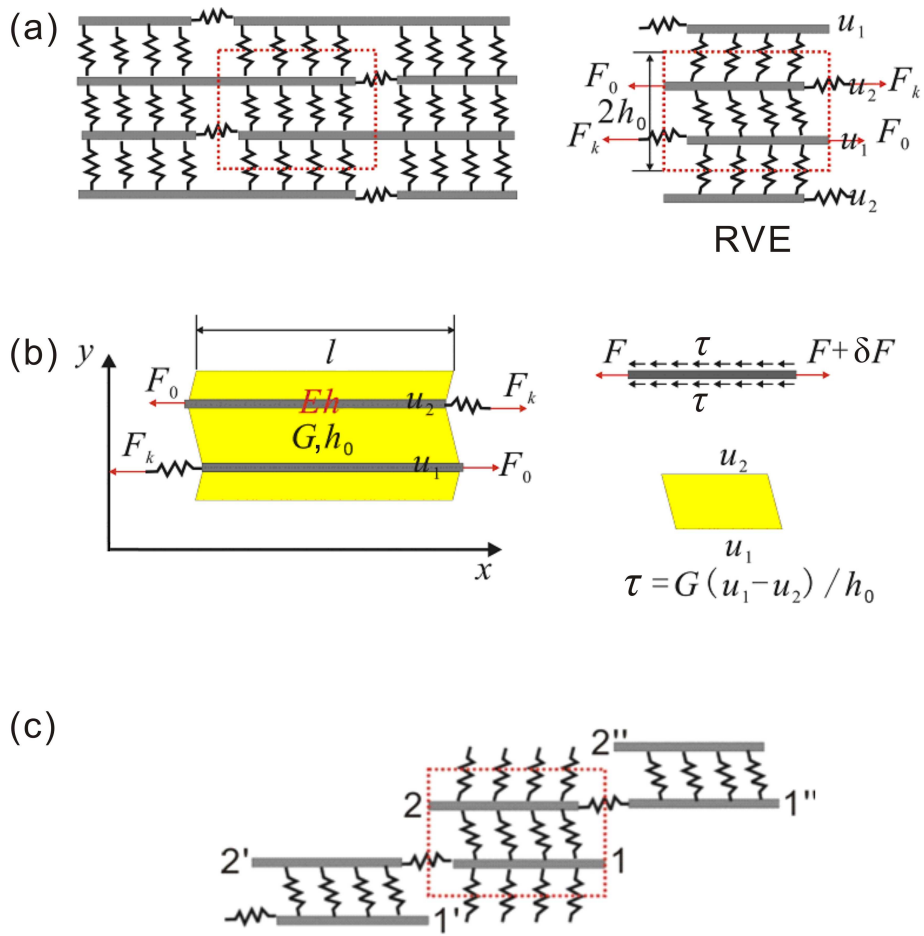

**Figure 2.** The schematic illustration of the analytical model. Subplot (a) shows the overall paper structure and representative volume element (RVE, as highlighted as the dash box) of the nano-composites. Subplot (b) shows the continuum representation of interlayer crosslinks (defined as a continuum in yellow) and parameters for geometry and mechanical properties, as used in the analytical model. Subplot (c) illustrates the load transfer between graphene sheets.

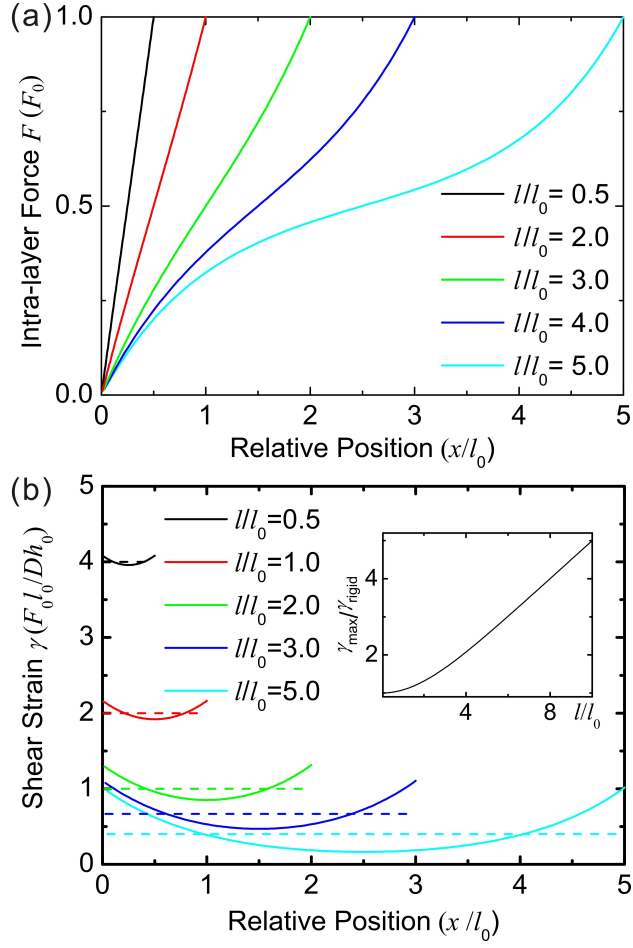

**Figure 3.** (a) The distribution of intralayer tensile forces $F(x)$ along the profile of the graphene sheets, showing linear behavior for small graphene sheets, and a flat region when the graphene sheet size increases. Correspondingly (b) gives similar plots for shear strain $\gamma(x)$ between adjacent graphene sheets. No intralayer crosslink is taken into account here. The inset shows the ratio between the maximum shear strain $\gamma_{max}$ in the DTS model and $\gamma_{rigid}$ in the rigid approximation for graphene sheets with different lengths.

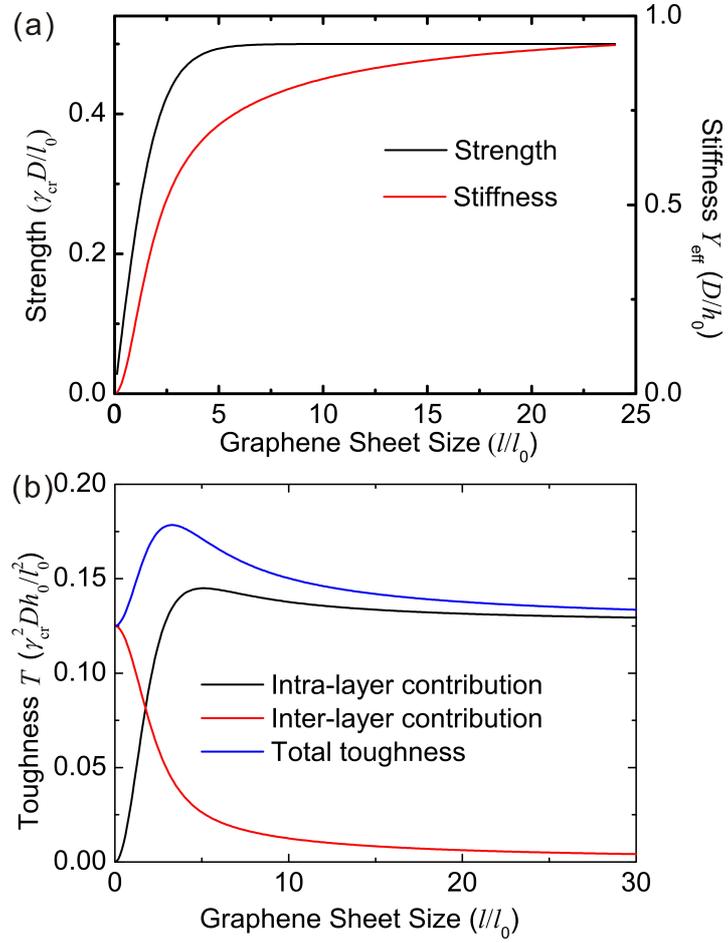

**Figure 4.** (a) The strength and stiffness of graphene-based nanocomposites, described by the analytical model developed in this work. The tensile strength converges quickly as the graphene sheet exceeds $5l_0$, and the stiffness also converges but at a larger graphene sheet size $l$. (b) The toughness of the nanocomposite, which reaches its maximum at $3.3l_0$. The toughness combines contributions from both intralayer and interlayer elasticity. While the first one (in-plane tension) also has a peak at $5l_0$ and dominates for large graphene sheets, the interlayer shear strain energy decreases as the graphene size increases.

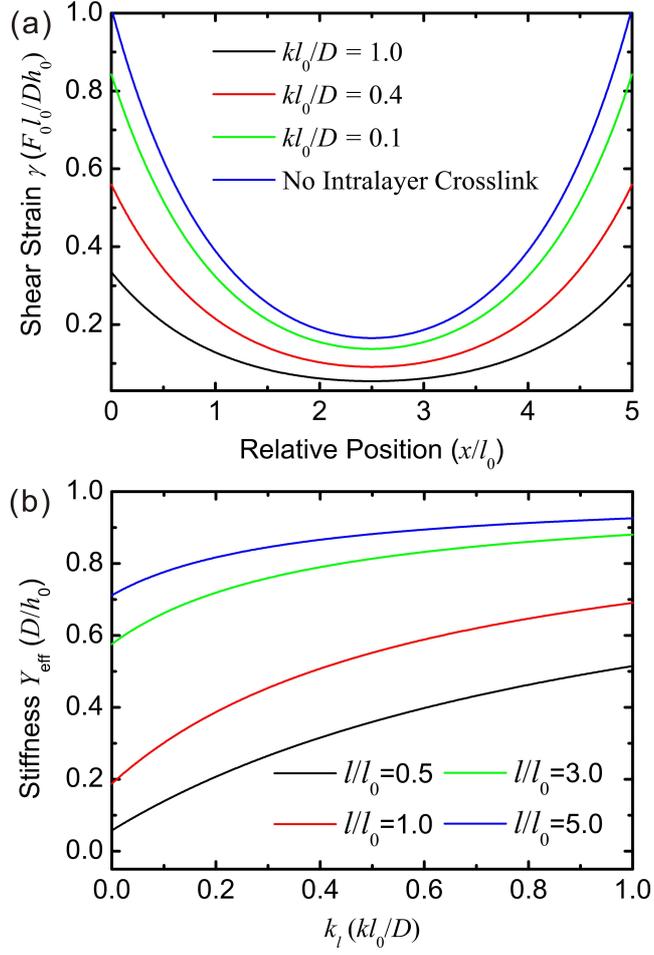

**Figure 5.** (a) Shear strain distribution in the interlayer crosslinks along the graphene sheet including the contribution of intralayer crosslinks, with spring constants $k$ varying between $0.1D/l_0$ and $10D/l_0$. (b) The relationship between the effective tensile modulus $Y_{eff}$ and the intralayer crosslink stiffness $k_l$

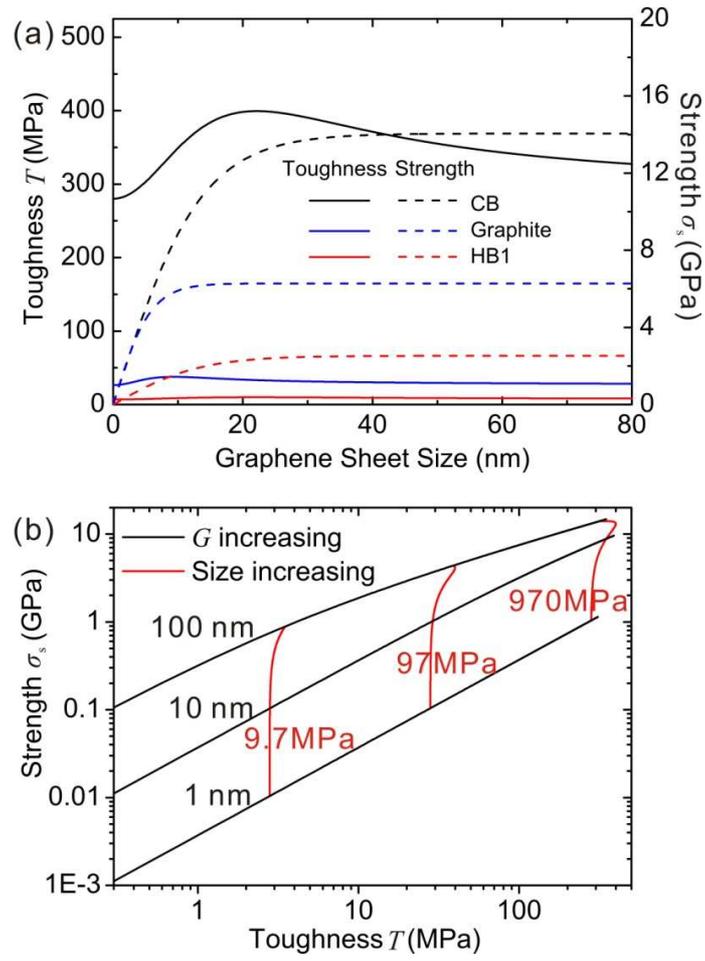

**Figure 6.** (a) Dependences of strength and toughness of the two representative crosslink types (coordinative bonds, CB and hydrogen bonds between epoxy and hydroxyl groups, HB1) and graphite. (b) The dependence of tensile strength and toughness on interlayer shear modulus $G$ and graphene sheet size $l$.